\documentclass[runningheads]{llncs}
\pdfoutput=1
\usepackage{booktabs}
\usepackage{geometry}
\usepackage{fancyhdr}
\usepackage{amsmath, amssymb}
\usepackage{graphicx}
\usepackage{caption}
\usepackage{subfigure}
\usepackage{hyperref}
\usepackage{multirow}
\usepackage{diagbox}
\usepackage{algpseudocode}
\usepackage{amssymb}
\usepackage[lined,boxed,commentsnumbered, ruled,vlined]{algorithm2e}
\usepackage{float}
\newfloat{algorithm2e}{t}{lop}
\usepackage{capt-of}
\usepackage{booktabs}
\usepackage{varwidth}

\begin{document}
	
	\title{Typical Snapshots Selection for Shortest Path Query in Dynamic Road Networks}

	\author{Mengxuan Zhang, Lei Li, Wen Hua, Xiaofang Zhou}
	\institute{{School of Information Technology and Electrical Engineering\\ The University of Queensland, Brisbane, Australia}\\
	\email{mengxuan.zhang@uqconnect.edu.au, \{l.li3, w.hua\}@uq.edu.au,zxf@itee.uq.edu.au}}
	\maketitle
	\pagestyle{empty}
	
	%
	%
	
	\begin{abstract}
		
	Finding the shortest paths in road network is an important query in our life nowadays, and various index structures are constructed to speed up the query answering. However, these indexes can hardly work in real-life scenario because the traffic condition changes dynamically, which makes the pathfinding slower than in the static environment. In order to speed up path query answering in the dynamic road network, we propose a framework to support these indexes. Firstly, we view the dynamic graph as a series of static snapshots. After that, we propose two kinds of methods to select the typical snapshots. The first kind is \textit{time-based} and it only considers the temporal information. The second category is the \textit{graph representation-based}, which considers more insights: \textit{edge-based} that captures the road continuity, and \textit{vertex-based} that reflects the region traffic fluctuation. Finally, we propose the \textit{snapshot matching} to find the most similar typical snapshot for the current traffic condition and use its index to answer the query directly. Extensive experiments on real-life road network and traffic conditions validate the effectiveness of our approach. 
	
	\end{abstract}
	
	
	\section{Introduction}
	\label{Introduction}
	
	Shortest path query is a fundamental operation in road network routing and navigation. A road network can be denoted as a directed graph $G(V, E)$ where $V$ is the set of road intersections, and $E \subseteq V\times V$ is the set of road segments. Each edge is associated with a numerical weight representing the length of a road segment or the time required to travel through. The road network is static if both the structure and the edge weights (i.e., $V$ and $E$) do not change over time. As for the real-life road network, the traffic condition changes almost all the time, we model the road network as a dynamic graph. Here we treat this dynamic graph as a series of snapshots, with each snapshot is static by itself but dynamic between each other, and answer the path queries using their corresponding snapshot graphs.
	
	The shortest path problem has been extensively studied and the approaches can be grouped into two categories depending on if an index is created or not. The \textit{index-free} methods like \textit{Dijkstra's}, \textit{A*} \cite{zhang2019batch,zhang2019Efficient}, and \textit{cache-based} \cite{thomsen2012effective}, find the path only with the graph information. Therefore, they can adapt to the dynamic environment by simply running on the new graph. But they suffer from low query efficiency or inaccurate results, so various \textit{index-based} methods \cite{geisberger2008contraction,ouyang2018hierarchy,samet2008scalable} have been proposed to speed up the query answering. However, these indexes all take time to construct, and the traffic condition may have already changed before their construction finishes. Therefore, there are two extreme cases to use index on a dynamic graph. The first one is building an index for each snapshot, which is not space efficient and has much redundant information. The other one is creating a big time-dependent index \cite{li2017minimal} for the entire time domain like \textit{TCH} \cite{batz2009time} and \textit{T2Hop} \cite{8731400}. However, their index sizes are huge and they essentially require the graph to be static from the perspective of "change". Therefore, we aim to seek a balance between the two extremes by identifying some typical snapshots from the dynamic graphs and only build indexes for them. Given queries in a specific current traffic condition, we first match the traffic condition to the most similar typical snapshot and use its index to answer the queries. When none of the existing snapshots is similar enough, we regard the current road network as a new snapshot and construct an index for it.
	
	However, it is unclear how to choose those typical snapshots and how to classify the current traffic condition. 
	We try to represent multiple similar snapshot graphs as one typical graph and then process queries in it at the cost of query accuracy. There are two lines of studies focus on graph similarity measurement. The first one is the \textit{graph edit distance} \cite{zhao2012efficient,gouda2016csi_ged,li2018efficient}. It uses the minimum edit operation number to transform one graph into another. The other one is the \textit{feature-based distance} \cite{chen2019efficient}, where the similarity is not measured on the graph directly but on the abstraction of a graph. Existing methods in both these two lines consider either attribute similarity or structural similarity. However, the road network is a special graph where the topological structure does not evolve frequently because the road construction and closure are not very common. In addition, the edges and vertices in the road network are not associated with labels. it is the edge weight or the speed that varies with time, and we suppose that the structure remains stable for the road network. Therefore, the existing graph similarity measurement can hardly solve our problem. However, we also represent the graph as a feature vector but use the edge weight vector and vertex vector, and we focus on the speed profile change rather than the change of the topological structure or the associated labels. To the best of our knowledge, it is the first time to specify the representation and the similarity measurement of the road network. Then multiple snapshots are clustered, the representative snapshot in each cluster can be selected as the typical snapshot. When there comes a query, we classify the current road network as the most similar typical snapshot and process the shortest path on it. 
	
	To support accurate clustering and classification of the snapshots, three challenges must be addressed. The first challenge is that the original representation of the road network encounters the high-dimension curse. Various dimension reduction techniques such as \textit{PCA} and \textit{LDA} are proposed, but they are general methods and none are specified to the road network. Here, we reduce the dimension by calculating the covariance of each edge first and filtering the edges whose values are smaller than the threshold. Also, we propose another two representations with much lower dimension: \textit{Edge-based} and \textit{Vertex-based}. The second challenge is how to incorporate the road network features like region property and road network continuity into the selection. The two proposed graph representation methods focus on different aspects of the network feature. The \textit{Edge-based} graph representation uses path to capture the continuity of the network. The \textit{Vertex-based} representation selects the "hot spots" (typical vertices) in networking by evaluating the fluctuation of the traffic condition around the vertices. The third challenge is how to choose the typical graph given the current traffic condition. When there comes a query, we convert the graph in the two proposed ways and then match the current graph with one typical graph by using the classification algorithm. The contributions in this work can be summarized as follows: 
	\begin{itemize}
		\item We formally study the problem of shortest path query in dynamic graphs.
		\item We propose two categories of the methods to select the typical snapshots: The \textit{time-based} approaches that choose the snapshots directly, and the \textit{graph-representation} approaches (\textit{edge-based} and \textit{vertex-based}) that consider the features like continuity, region condition. 
		\item We present how to do the graph clustering and classification in our \textit{graph-representation}.
		\item We conduct extensive evaluations using a large real-world road network and traffic condition. The experimental results verify the effectiveness of our approaches.
	\end{itemize}
	
	The remaining of this paper is organized as follows: We first discuss the current literature of the pathfinding and graph similarity measurement in Section \ref{RelatedWork}. Section \ref{Preliminary} introduces some common notions and defines the sub-problems: \textit{Typical Snapshots Selection} and \textit{Snapshot Matching}. For the first sub-problem, we propose two time-based approaches in Section \ref{sec:Baseline}, and present two graph representation-based methods in Section \ref{sec:GraphRepresentation}. The actual typical snapshot selection and the second matching sub-problem are discussed in Section \ref{sec:SimilarityAndCluster}. Evaluations of the proposed methods in a real-life dynamic road network are presented in Section \ref{ExperimentNew}. Finally, Section \ref{Conclusion} concludes the paper.
	
	\section{Related Work}
	\label{RelatedWork}
	
	\subsection{Shortest Path Algorithm}
	
	In the past decades, various techniques have been proposed for the shortest path calculation in road networks. The fundamental shortest path algorithms are \textit{Dijkstra's} and \textit{A*} algorithms. The \textit{Dijkstra's} is inefficient as it needs to traverse the entire network for the shortest path search. And the \textit{A*} improve the query efficiency by directing the traversal towards the destination with the help of the heuristic distance. Then there comes a line of research that accelerates the query answering by pre-calculating the index. Particularly, algorithms such as \textit{Contraction Hierarchy} \cite{geisberger2008contraction} prune the search space by referring to information stored in the index. Other algorithms like \textit{2-Hop Labeling} \cite{ouyang2018hierarchy} and \textit{SILC} framework \cite{samet2008scalable}, attempt to materialize all the pairwise shortest path results in a concise or compressed manner such that a given shortest path query can be answered directly via a simple table-lookup or join. These index-based algorithms are usually efficient for query answering so as to return the query result within microseconds, but the major premise behind them is that the road network should be static. Since the index construction is often time-consuming and the road network evolves almost all the time, the rebuilt index cannot always fit the current refreshed network condition. Therefore, these algorithms do not adapt well to the dynamic environment.
	
	Another line of research attempts to process the query in dynamic environment. It uses functions to describe the road condition directly \cite{li2017minimal,li2018go}. However, the complexity of finding the fastest path is $\varOmega(T(|V|\log |V|+|E|))$, where $T$ is a large number related to the function. This complexity lower bound determines it is much slower to find a fastest path compared with the static environment. To further speed up the query efficiency, time-dependent indexes like \textit{TCH} \cite{batz2010time} and \textit{T2Hop} \cite{8731400} are proposed. However, these time-dependent algorithms essentially view the dynamic environment statically, because their time-dependent functions are stable. Any change of their function would result in the failure of the existing index and have to endure a time-consuming reconstruction process. Therefore, some works drop the time-dependent functions and run the query in the dynamic graph directly. Because it is hard to build an index for the dynamic graph, shared computation \cite{zhang2019batch,zhang2019Efficient,li2020fast} is introduced to improve the query efficiency. Nevertheless, their efficiency is still not comparable with index-based approaches. In this work, we aim to bring the index back to the dynamic environment with the help of snapshots.

	\subsection{Graph Similarity Measurement}
	\label{subsec:RelatedWork_GraphSimilarity}
	The graph distance can be measured mainly in two ways: \textit{graph edit distance} \cite{zhao2012efficient,li2018efficient} and \textit{feature-based distance} \cite{yan2005substructure,zhou2009graph}. \textit{Graph edit distance} has been widely accepted for the graph similarity measurement, and two graphs whose distance is less than a similarity threshold is considered to be similar. It is a metric which can be used in various graphs such as directed or undirected, labeled or unlabeled, as well as single or multi- graphs. The distance is calculated as the minimal steps of graph edit operations including the insertion, deletion, or alteration of vertex or edge to transform one graph to another. In this way, this method can reflect the topological differences between graphs. However, it is not applicable to our problem because the topological structure of the road network does not change often and is supposed to be static here. For the \textit{feature-based distance}, most of the existing works focus on the structure-based, attributed-based or structural/attribute distance. In \cite{yan2005substructure}, some labeled edges are selected as the features and one graph is represented as a feature vector where each dimension indicates the existence or the frequency of the corresponding edge. The neighborhood random walk model is proposed to combine the structural closeness and attribute similarity for the graph clustering \cite{zhou2009graph}. However, in our scenario, the edges are associated with the length or travel time rather than the labels, and the structure is supposed to be static as mentioned above. In this work, we aim at distinguishing multiple snapshots by their speed profile, and we need to measure the graph similarity from the combination of edge weights and graph structure. Therefore, the existing graph similarity measurement is difficult to be applied here. 
	
	\section{Problem Definition}
	\label{Preliminary}
	\begin{definition}
		\label{def:RoadNetwork}
		\textbf{(Road network)}. Road network is formalized as a dynamic weighted graph $G_D(V, E, W(T))$, where the vertex $v\in V$ (resp. edge $e\in E$) denoting road intersection (resp. segment) and the edge weight $w(e,t)\to \mathbb{R},t\in T$ can change with time.  
	\end{definition}
	
	If we take a snapshot of the dynamic graph at some time point $t$, then each edge on the snapshot $g_i=G_S(V, E, W(t_i))$ is associated with only one weight value. Suppose the road traffic condition is constant around a small time interval of the snapshot, then the dynamic graph can be treated as the set of multiple snapshots with the timestamp, that is $G_D=\{g_i=G_S(V, E, W(t_i))| t_i\in T, \bigcup t_i=T\}$. 
	
	We focus on the shortest path query in the dynamic road network. Given $k$ typical snapshots with their corresponding indexes, and the shortest path queries in the current road network, we try to match the current graph to the most similar typical graph and use its index for the query answering. Therefore, two sub-problems \textit{typical snapshot selection} and \textit{snapshot matching} appear and they are defined as followed.
	
	\begin{definition}
		\label{def:sub1}
		\textbf{(Sub-Problem 1: Typical Snapshots Selection)}. Given multiple snapshots $G=\{g_0, g_1,\dots, g_{n-1}\}$ of a road network, \textit{typical snapshots selection} puts them into $k$ ($k<n$) clusters such that the snapshots in the same cluster are similar and those in different clusters are dissimilar. One representative snapshot is taken from each cluster as the typical snapshot.
	\end{definition}
	
	\begin{definition}
		\label{def:sub2}
		\textbf{(Sub-Problem 2: Snapshot Matching)}. Given one snapshot $g_i$ and $k$ typical snapshots $G_T=\{g_1, g_2,\dots,g_k\}$ of the road network, where $g_i$ is not necessary in $G_T$, \textit{snapshot matching} captures captures the snapshot $g^*$ that is the most similar with $g_i$ from $G_T$. 
	\end{definition}

	Apparently, both of the two sub-problems need the graph similarity measurement. We measure the similarity of road networks by first abstracting the features and then use the distance between the feature vectors as the graph similarity. 
	
	To evaluate the difference or quality between the selected typical graph $g_i$ and the actual graph $g^*$, we need an error measurement. We compute the traveling time $l_i(p)$ and $l^*(p)$ by using the edges of $g_i$ and $g^*$ for each $p\in P$. The error $p$ is $error(p)=|l_i(p)-l^*(p)|/|l^*(p)|$, and the error between $g_i$ and $g^*$ is $error(g_i,g^*)=\dfrac{\sum error(p)}{|P|}$. Because the trajectories are collected from the taxi, this measurement focuses more on the actual impact on real-life traveling.
	
	\section{Time-Based Typical Snapshot Selection}
	\label{sec:Baseline}
	The dynamic road network can be viewed as a time series of snapshots. Because the traffic on road network changes incrementally in real life and several continuous snapshots can be approximately the same. Based on the observation, we can select the typical snapshots by sampling on the time dimension. In the following, we present two baseline selection methods: \textit{uniform sampling} and \textit{non-uniform sampling}.  
	\vspace*{-5mm}
	\subsection{Uniform Sampling}
	\label{subsec:Baseline_Uniform}
	Suppose the total snapshot number is $n$. The uniform sampling method selects the snapshots with the same step $x$ starting from the $y^{th}$ snapshot ($y<x$). In other words, $G_T=\{g_i|i=y+kx, 0\leq i< m\}$. When $x=1$, all $n$ snapshots are selected, and its error is 0; when $x=2$, every odd or even snapshot is selected, and it has some error; when $x\leq \dfrac{n}{2}$, only one snapshot is selected, and it has the largest error. The number of the typical graph is $k=\lfloor m/x\rfloor$. Obviously, the error could be inversely proportional to the typical snapshot number $k$, and we test the performance of different $k$. This method can control the number of snapshots, but it cannot guarantee the worst case error. The time complexity is $O(n)$. 
	\vspace*{-5mm}
	\subsection{Non-Uniform Sampling}
	\label{subsec:Baseline_NonUniform}
	The change rate of traffic conditions differs in each time period. For example, the road network is almost the same from midnight to the early morning because little traffic appears on road. But it can change dramatically during peak hours. Therefore, we select the typical snapshots non-uniformly according to how the traffic changes by time, which can be captured by the path-based error. 
	
	The sampling works in a sliding window fashion. First of all, an error threshold $\epsilon$ is set. After that, we visit the snapshots in the increasing order of timestamps and put the current visiting snapshot $g_i$ into the current window $G'$. For each $g_j\in G'$, we compute its error $error'(g_j)=max(error(g_j,g_i))$, where $g_i\in G'$ and $g_j \neq g_i$. Then the $g_j$ with the minimum $error'(g_j)$ is selected as the typical graph of the current window. If $error'(g_j)\leq \epsilon$, the windows keep expanding and test the next snapshot. Otherwise, a typical graph selected for the previous windows and a new window is created with $g_i$ as the first snapshot. This procedure runs on until all the snapshots are visited. Apparently, this method can control the worst error, but it cannot determine the number of typical snapshots. The time complexity is $O(n^2)$.
	
	\section{Graph Representation-Based Selection}
	\label{sec:GraphRepresentation}
	
	\subsection{Edge-Based Representation}
	\label{subsec:Graph_Edge-based}
	
	Suppose the road network structure does not change, which means $V$ and $E$ is steady, then only the weight vectors differ for multiple graphs. Therefore, in our first type of representation, we use the \textit{weight vector} and the \textit{delta weight vector} to denote one graph.
	
	\subsubsection{Single Edge Representation}
	
	If we denote one graph as the weight vector, then one snapshot $G_i$ is directly represented as $W_i=[e_1,e_2,\dots,e_{|E|}]$, which contains every edge's weight. And its variant is the delta weight vector, that is we can express one snapshot $g_i$ as $\delta(W_i)$, where $W_0$ is the weight vector of $g_0$ (treated as referenced graph), and $\delta(W_i)=W_i-W_0=[\delta e_1,\delta e_2,\dots,\delta e_{|E|}]$. Because both of $W_i$ and $\delta(W_i)$ has the same dimension number of $|E|$, which could be hundreds of thousands in real-life and suffers from the curse of dimension, we have to reduce the dimension number before computing the similarity.
	
	The first approach of dimension reduction we apply is \textit{PCA (Principal Component Analysis)}. However, it is a general dimension reduction algorithm and does not perform well in our scenario. In the road network, it is those edges that change dramatically over that distinguish a typical snapshot. Then we use the coefficient of variation $cv$ (standard deviation divided by the mean) of each edge to measure how various an edge is and use a threshold to identify those various ones to construct a lower-dimensional weight vector. 
	\vspace*{-6mm}
	\subsubsection{Aggregated Edge Representation}
	The weight vector shows the weight of every edge in a graph, but it loses the information of the connection and continuity of road segments. 
	Usually, it is the continuous road segments in some areas that are busy or congested rather than the individual road segments or all the road segments in one area. Therefore, we try to use the aggregated edge length of multiple paths to represent one graph and we call these paths as \textit{typical paths}. 
	
	Each path $p$ is a sequence of connected edges with $p=[e_0,e_1,\dots,e_n]$ and the length of a path is $d(p)=\sum_{i=0}^{n}w(e_i)$. Suppose there are $k$ typical paths, then one snapshot is represented as $g^{AE}=[d(p_0),d(p_1),\dots,d(p_{k-1})]$.
	
	To better represent a graph, the typical paths set should meet the following conditions: 1) The coverage of typical paths should be as large as possible to represent the graph completely; 2) The similarity between typical paths should be small to avoid the redundant representation; 3) The length (calculated as the total time passing through) of the same path should vary greatly so as to differ multiple snapshots . And according to the observation of traffic in daily life that the congested road segments are usually within local areas, such as the discontinuous red or yellow segments along one long path, we set the minimum static length $l_{min}$ and the static maximum length $l_{max}$ (calculated as the total length) of the candidate typical paths as $2km$ and $3km$, respectively. 
	
\begin{figure}[htbp]
		\centering
		\begin{minipage}[t]{0.45\textwidth}
			\centering
			\includegraphics[width=5cm]{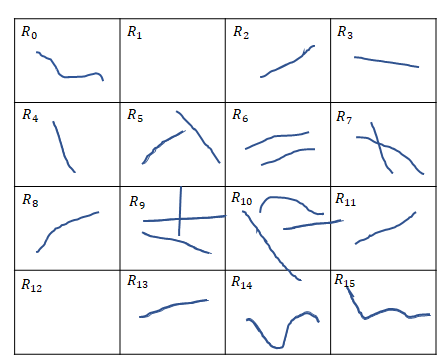}
			\caption{Typical paths distribution}
			\label{fig:path1}
		\end{minipage}
		\begin{minipage}[t]{0.45\textwidth}
			\centering
			\includegraphics[width=5cm]{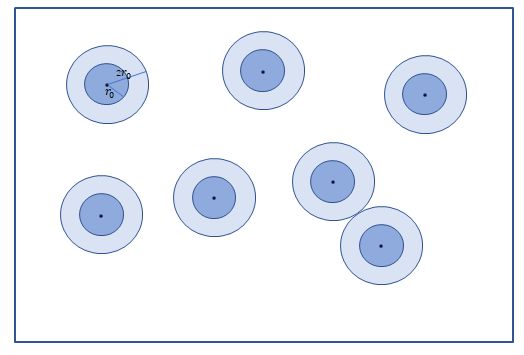}
			\caption{Typical Vertex Selection}
			\label{fig:vertex1}
		\end{minipage}
	\end{figure}

	To meet the first condition of path selection, we partition the graph evenly into $4\times 4$ regions $R=\{R_0,R_1,\dots,R_{15}\}$. An example of the selected paths in each region is shown in Figure \ref{fig:path1}. The selected paths number $pnum_i$ in region $R_i$ is proportional to the vertices number in it with $pnum_i=pnum\times|V_i|/|V|$, where $pnum$ is the typical paths total number and $|V_i|$ is the vertices number in $R_i$. The paths generated at this step are the candidates. To meet the second condition, we compute the similarity between typical paths in a region and remove one of those that are larger than a threshold. The similarity here is the Jaccard Coefficient ($\dfrac{|p_i\cap p_j|}{|p_i\cup p_j|}$) over the edges. In the following, we present different ways to select the candidate typical paths.
	
	\textbf{Random Selection} The simplest way is to randomly select $pnum_i$ paths in each region $R_i$ with path length between $l_{min}$ and $l_{max}$. First of all, a length threshold $\eta$ is determined randomly. After that, a starting edge is selected randomly, and we choose one of its neighbors randomly. The path keeps growing until the length is larger than $\eta$. Repeated edge is avoided for better coverage. 
	
	\textbf{Edge-Constrained Selection} To increase the representativeness of the typical paths, we select those paths whose edges' coefficient of variations is no less than a threshold $thresh_{cv}$. However, it is inefficient to generate and validate candidates forwardly, so we do the coefficient of variation filtering first and construct a sub-graph only with the highly changing edges. After that, the candidates are generated on this sub-graph instead of the original graph. 
	
	\textbf{Edge-Greedy Selection} The random selection ignores the weight variation totally so it suffers from generating the qualified candidates repeatedly, while the edge-constraint selection is limited to a small sub-graph so it faces the headache of high similarity between the candidates. Therefore, we propose a greedy method to generate the candidates considering both the weight variation and path distinction. This approach also runs on the original graph and select the starting edge randomly. As for growing the candidate path, the next selected out-edge is the one with the largest $cv$. To maintain the effectiveness of the path, a smaller threshold $thresh_{cv}$ is applied to validate the edge. 
	\vspace*{-10mm}
	\vspace*{-12mm}
	\subsection{Vertex-Based Representation}
	\label{subsec:Graph_Vertex-based}
	
	In real life, there always exist some temporal hot spots in the road network, such as the inevitable road intersection during rush hour, the scenic spots on weekend, and the business area after work. Meanwhile, the traffic conditions in other ``cold" areas stay normal at the same time. Then how about detecting these ``key" vertices (called \textit{typical vertices}) and use the aggregation of traffic conditions around them to represent the traffic condition on the whole road network? Consequently, two problems need to be solved: 1) how to find the typical vertices? 2) how to use the typical vertices to represent one snapshot for the similarity measurement among different snapshots?
	
	\subsubsection{Graph Representation}
	Suppose these typical vertices $V_T=\{v_i\},v_i\in V$ in a road network have already been known in advance. Inspired by the \textit{tree-based q-gram approach} for graph similarity join problem \cite{wang2010efficiently,zhao2012efficient}, we represent the traffic condition around each typical vertex $v_i$ as the set of vertices that can be reached from $v_i$ in a breath-first-search within a fixed time period (for example, 2 minutes). If a driver arrives at a hot spot, he or she is likely to be blocked by the traffic flow and could pass through fewer road intersections within the time period. Usually, the smoother the traffic condition around $v_k$ in $g_i$ is, the larger the value of $|S_{ik}|$ will be, and vice versa. Obviously we cannot learn much from the absolute value of $|S_{ik}|$, and we care more about the congestion than the smoothness of the traffic. We define the block coefficient of a vertex $v_k$ in $g_i$ as
	\begin{equation}
	b(v_{ik})=max\{|S_{1k}|,|S_{2k|},\dots,|S_{nk}|\}/|S_{ik}|
	\end{equation}
	where $max\{|S_{1k}|,|S_{2k|},\dots,|S_{nk}|\}$ represents the maximum reachable vertex number from $v_k$ among multiple snapshots, and it reflects the non-block traffic condition around $v_k$ in other words. The larger the block coefficient $b(v_{ik})$, the more congested around $v_k$ at time period $t_i$. 
	
	In the first type of vertex-based representation, we denote one snapshot as the block coefficient of the typical vertices (called \textit{vertex-bc representation}), that is $g_i=[b(v_{i0}),b(v_{i1}),\dots,b(v_{i|V_T|})]$. We can also represent one snapshot as the vertex set of typical vertices (called \textit{vertex-set representation}), that is $g_i=[S_{i1}, S_{i2}, \dots, S_{i|V_T|}]$ with $S_{ik}$ denoting the vertex set reached from $v_k$ within $t_0$ time in $g_i$. And the reachable vertex set from the vertex $v_k$ in $n$ snapshots can be denoted as $[S_{1k},S_{2k},\dots,S_{nk}]$. 
	
	\subsubsection{Typical Vertices Selection}
	
	The difference between the hot spots and the "cold" vertices is that the traffic condition fluctuates more dramatically around the hot spots. Hence, we define the traffic fluctuation $f(v_k)$ of $v_k$ as the coefficient of variation of the block coefficient:
	\begin{equation}
	f(v_k)=\frac{\sigma\{b(v_{1k}),b(v_{2k}),\dots,b(v_{nk})\}}{\mu\{b(v_{1k}),b(v_{2k}),\dots,b(v_{nk})\}}
	\end{equation} 
	where $\sigma$ and $\mu$ denotes the standard deviation and the mean of $\{b(v_{1k}),b(v_{2k}),\dots,b(v_{nk})\}$..
	
	To select the typical vertices, we visit the vertices in decreasing order of their traffic fluctuation and choose the top $|V_T|$ vertices. Besides, during the selection, we exclude the vertices that are close to the selected typical vertices because they are likely to capture the traffic condition of the overlapped local area or have the similar traffic fluctuation pattern. 
	
	Specifically, We first compute the vertex set $S_{ik}$ for each vertex on each snapshot using \textit{BFS}. However, the search does not stop at $r$ but at $2r$ and also generates a larger coverage set $S'_{ik}$. $S_{ik}$ is used to compute the block coefficient $b(v_{ik})$ and traffic fluctuation $f(v_k)$, while $S'_{ik}$ is used to avoid the typical vertices being too close to each other.  As shown in Figure \ref{fig:vertex1}, the vertex coverage set of the selected vertices have no intersection with each other. The procedure stops when $k$ typical vertices are selected. The time complexity is $O(|V||G_D|\times BFS(2r) + |V|\log |V|)$. Because the complexity of the \textit{BFS} is dependent on a small radius $2r$, we use $BFS(2r)$ to denote its complexity. 
	
	\subsection{Graph Clustering and Snapshot Matching}
	\label{sec:SimilarityAndCluster}
	In this section, we discuss the methods to solve the two sub-problems. The previous section introduced two types of graph representations, and we present how to utilize them to select the typical snapshots. Section \ref{subsec:SimilarityAndCluster_Clustering} presents how the typical snapshots are determined and \ref{subsec:SimilarityAndCluster_Classification} solves the \textit{snapshot matching} sub-problem with graph classification.
	
	\subsubsection{Graph Clustering}
	\label{subsec:SimilarityAndCluster_Clustering}
	Because the graphs are represented as low-dimensional vectors, we can utilize general clustering methods to put similar ones together. However, methods that tend to cluster arbitrary shapes like \textit{DBSCAN} \cite{ester1996density,gan2015dbscan} are not suitable for tasks like this because of their errors are not guaranteed. Therefore, we use two types of methods: \textit{adaptive K-means based clustering} and \textit{agglomerative hierarchical clustering}\cite{defays1977efficient} that have a distance threshold to do the clustering.

	\subsubsection{Graph Classification}
	\label{subsec:SimilarityAndCluster_Classification}
	When the traffic condition changes, we can receive a new snapshot $g'$. First of all, $g'$ is converted into one of the graph representations. After that, it is compared with the existing typical snapshots and obtains the most similar one $g^*$. If the similarity between satisfies the threshold, we use the index of $g^*$ directly to answer the path queries. Otherwise, $g'$ is considered as a new typical snapshot, and a new index is also built for it.
	
	\section{Experiments}
	\label{ExperimentNew}
	In this section, we experimentally evaluate the performance (in terms of the accuracy and efficiency) of the proposed \textit{typical snapshot selection} and \textit{snapshot matching} approaches using the real-life road network with real traffic condition.
	\vspace*{-8mm}
	\begin{figure}[htbp]
		\begin{minipage}[t]{1\textwidth}
			\centering 
			\subfigure[Single-Edge Representation] { 
				\includegraphics[width=0.45\textwidth]{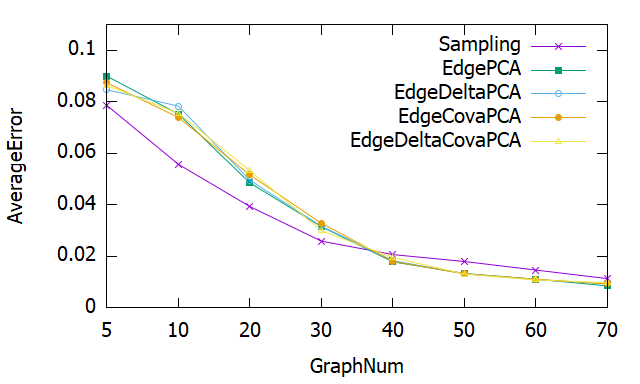}
				\label{fig:edge1} 
			} 
			\subfigure[Aggregated Edge Representation] { 
				\includegraphics[width=0.45\textwidth]{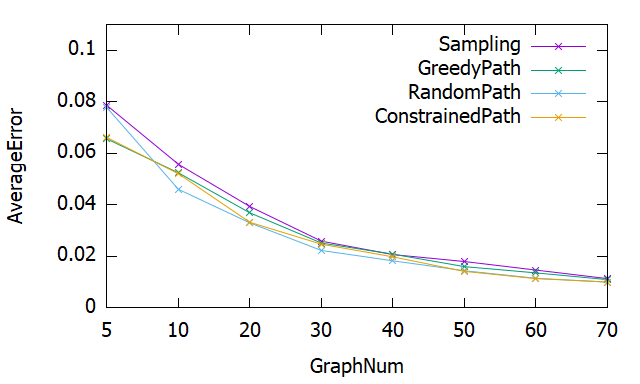} 
				\label{fig:edge2} 
			} 
			\caption{Performance of Edge-Based Representation}
			\label{fig:edge} 
		\end{minipage}
	\end{figure}
	\vspace*{-15mm}
	\subsection{Experimental Setup}
	We execute the experiments on the Beijing road network with 31,2350 vertices and 40,3228 edges. Currently, there are 288 snapshots sampled every 5 minutes from the traffic condition in 1$^{st}$ April 2015. These snapshots are obtained from the taxi trajectories collected during that day. The original trajectory dataset contains 532,868 trajectories and 17,698,668 GPS points. We follow the same process of \cite{li2018go} to generate the speed profile. 
	
	
	
	
	All the algorithms are implemented in C++, compiled with full optimizations, and tested on a Dell R730 PowerEdge Rack Mount Server which has two Xeon E5-2630 2.2GHz (each has 10 cores and 20 threads) and 378G memory. The data are stored on a 12 $\times$ 4TB Raid-50 disk.
	
	\subsection{Typical Snapshot Selection}
	\label{subsec:Experiment_TSS}
	\subsubsection{Edge-Based Representation}
	Figure \ref{fig:edge1} shows the result of the \textit{single-edge representation} test. These snapshots are clustered by \textit{K-means}. The edge vector and the edge delta vector are denoted as \textit{EdgePCA} and \textit{EdgeDeltaPCA}, with the PCA dimension reduction. And we use \textit{EdgeCovaPCA} and \textit{EdgeDeltaCovaPCA} to denote the combine of coefficient of variant and PCA dimension reduction. The performance of \textit{single-edge presentation} is better than the \textit{uniform sampling} method only when the typical snapshot number is over 40. Because in this kind of representation, we only consider the weight of edge and ignore the connectivity of edges and the underlying topological structure. Although the graph structure stays the same for each snapshot, it has a great impact on the location of the shortest path.
	
	Figure \ref{fig:edge2} shows the performance of the \textit{aggregated-edge representation}. The \textit{random/edge-constrained/edge-greedy} path selections are denoted as \textit{RandomPath}, \textit{ConstrainedPath}, and \textit{GreedyPath}. It can be seen that the aggregated-edge representation performs slightly better than the uniform sampling method. And the performance of these three variants is pretty much the same, which indicates that the typical paths are still not enough to represent the snapshot. But since the edge connectivity is considered in this representation, it performs better than the singe-edge representation (all the lines are below the \textit{sampling}, while the half of the single-edge's lines are above the \textit{sampling}).
	
	\subsubsection{Vertex-Based Representation}
\begin{figure}[htbp]
		\begin{minipage}[t]{1\textwidth}
			\centering 
			\subfigure[50 Typical Vertices] { 
				\includegraphics[width=0.45\textwidth]{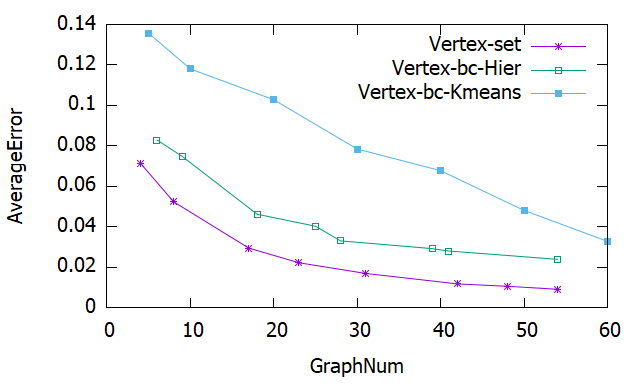}
				\label{vertex50} 
			} 
			\subfigure[100 Typical Vertices] { 
				\includegraphics[width=0.45\textwidth]{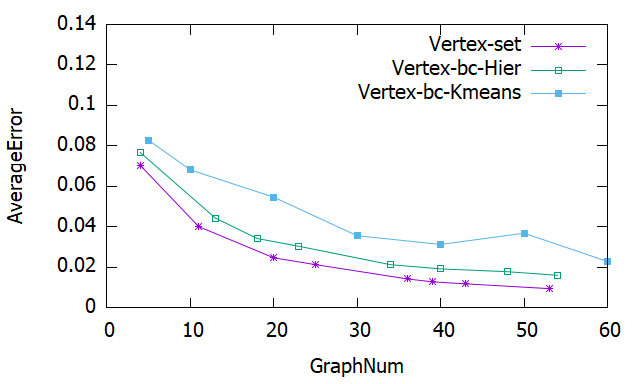} 
				\label{vertex100} 
			} 
			\subfigure[150 Typical Vertices] { 
				\includegraphics[width=0.45\textwidth]{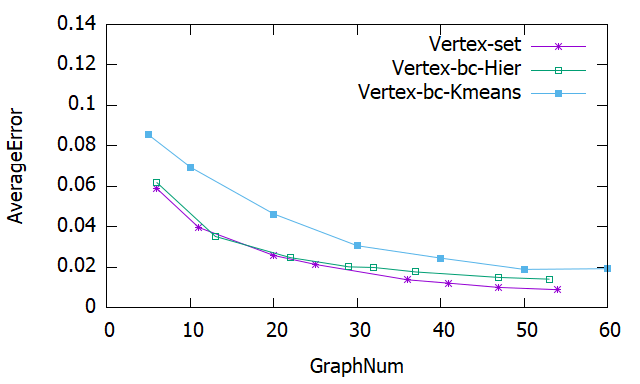} 
				\label{vertex150} 
			} 
			\subfigure[200 Typical Vertices] { 
				\includegraphics[width=0.45\textwidth]{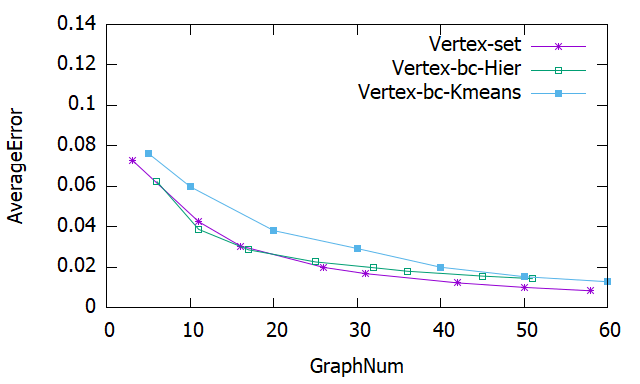} 
				\label{vertex200} 
			} 
			\caption{Performance of Vertex-Based Representation.}
			\label{fig:vertexset} 
		\end{minipage}
	\end{figure}

	For the \textit{vertex-set} representation, we cluster the snapshots by \textit{hierarchical clustering} and the testing performance is named as \textit{vertex-set}. And we use both the \textit{K-means} and \textit{Hierarchical Clustering} in the \textit{vertex-bc} representation and the results are denoted as \textit{vertex-bc-Hier} and \textit{vertex-bc-Kmeans} as shown in Figure \ref{fig:vertexset}. 
	\vspace*{-8mm}
	\vspace*{-8mm}
	In terms of vertex-based representation, it can be seen that the shortest path error of \textit{vertex-set} representation is always smaller than that of the \textit{vertex-bc} representation regardless of the clustering methods. In \textit{vertex-bc} representation, we consider both the reachable vertex number and the vertex set distribution overlapping, which is proved reasonable in these experimental results. In terms of typical vertex number, the error decreases distinctly when the typical vertex number rise from 50 to 150. It makes sense because more typical vertices can represent the snapshot and show the traffic characteristics more completely so as to generate more accurate clustering results. When the typical vertex number increase from 150 to 200, the errors are almost the same for all three methods. This indicates that taking less than 150 typical vertices is enough to represent the snapshot. What's more, fewer typical vertices is good for improving the snapshot matching efficiency. It is interesting to find that the performance of selecting 50 typical vertices is almost the same as that of 200 typical vertices, which again shows the superiority of vertex-set representation.
	
	\subsubsection{Time-Based Selection}
	In this section, we compare the performance of the \textit{time-based} methods and the \textit{graph representation-based} methods. For the representation method results, we choose \textit{ConstrainedPath} from the \textit{edge-based}, \textit{vertex-set} from the \textit{vertex-based} because they are the best of their own categories. The result is shown in Figure \ref{fig:time}. The worst method is the \textit{uniform sampling}, followed by the \textit{non-uniform sampling}. The three \textit{graph representation-based} methods are all better than the \textit{time-based} methods. Specifically, \textit{vertex-based} is better than \textit{edge-based}.
	\vspace*{-5mm}
	\begin{table}[ht]
		\begin{varwidth}[b]{0.5\linewidth}
			\centering
			\begin{tabular}{|c|c|c|}
				\hline
				\textbf{}              & \textbf{Edge-Based}   & \textbf{Vertex-Based}     \\ \hline
				\textbf{Graph Convert} & 6.012$\times 10^{-6}$ & $4.121\times 10^{-4}$     \\ \hline
				\textbf{Similarity}    & $k\times 10^{-8}$     & $k\times4.9\times10^{-5}$ \\ \hline
			\end{tabular}\hfill
			\caption{Snapshot Matching Running Time (sec)}
			\label{table:SM}
		\end{varwidth}%
		\hfill
		\begin{minipage}[b]{0.5\linewidth}
			\centering
			\includegraphics[width=70mm]{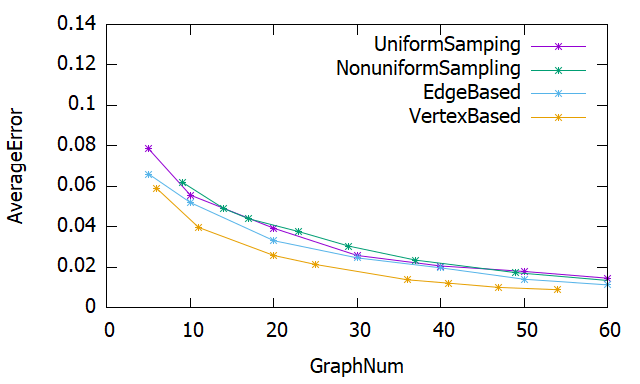}
			\captionof{figure}{Performance comparison of four methods}
			\label{fig:time}
		\end{minipage}
	\end{table}
	
	\vspace*{-15mm}
	\subsection{Snapshot Matching}
	\label{subsec:Experiment_SM}
	In this section, we evaluate the running time of the \textit{snapshot matching} procedure. Because the graph representation-based methods have higher accuracy than the time-based methods, we only show their results. The matching time is made up of the \textit{representation time}, which convert the current snapshot into one of the representations, and the \textit{similarity computation time}, which compares with the existing typical snapshots and finds the most similar one. Specifically, \textit{matching time}$=t_{r}+k\times t_{Similarity}$, where $k$ is the number typical snapshots. 
	
	The result is shown in Table \ref{table:SM}. The \textit{Edge-based} is the fastest to run because it only needs edge weight concatenation. The \textit{Vertex-based} is slower because it has to run hundreds of \textit{Dijkstra's} to collect the vertex set. Nevertheless, all of these methods can finish in one second, and the matching process like this only needs to run once when the traffic condition changes. 
	
	\section{Conclusion}
	\label{Conclusion}
	In this paper, we study the problem of supporting the index-based shortest path query answering in the dynamic road network. Because of the dynamic nature of the real-life traffic condition, none of the existing index structures can adapt to the real dynamic environment. On the other hand, although the traffic condition changes over time, it does not change dramatically in a short period. Therefore, we view the dynamic road network as a series of snapshots and only build the indexes on the typical ones. The first problem is how to determine if one snapshot is typical or not. We propose two sets of \textit{time-based} and \textit{graph representation-based} approaches to deal with it. After that, when facing a new traffic condition snapshot, we use the \textit{snapshot matching} to find the most similar typical snapshot, and use its index to answer the path queries. Our extensive experiments use the real-life road network, traffic condition to validate the effectiveness of our methods.	
	
	
\bibliographystyle{IEEEtran}
	\bibliography{references}

\begin{thebibliography}{10}
\providecommand{\url}[1]{#1}
\csname url@samestyle\endcsname
\providecommand{\newblock}{\relax}
\providecommand{\bibinfo}[2]{#2}
\providecommand{\BIBentrySTDinterwordspacing}{\spaceskip=0pt\relax}
\providecommand{\BIBentryALTinterwordstretchfactor}{4}
\providecommand{\BIBentryALTinterwordspacing}{\spaceskip=\fontdimen2\font plus
\BIBentryALTinterwordstretchfactor\fontdimen3\font minus
  \fontdimen4\font\relax}
\providecommand{\BIBforeignlanguage}[2]{{%
\expandafter\ifx\csname l@#1\endcsname\relax
\typeout{** WARNING: IEEEtran.bst: No hyphenation pattern has been}%
\typeout{** loaded for the language `#1'. Using the pattern for}%
\typeout{** the default language instead.}%
\else
\language=\csname l@#1\endcsname
\fi
#2}}
\providecommand{\BIBdecl}{\relax}
\BIBdecl

\bibitem{zhang2019batch}
M.~Zhang, L.~Li, W.~Hua, and X.~Zhou, ``Batch processing of shortest path
  queries in road networks,'' in \emph{Australasian Database Conference}.\hskip
  1em plus 0.5em minus 0.4em\relax Springer, 2019, pp. 3--16.

\bibitem{zhang2019Efficient}
------, ``Efficient batch processing of shortest path queries in road
  networks,'' in \emph{2019 20th IEEE International Conference on Mobile Data
  Management (MDM)}.\hskip 1em plus 0.5em minus 0.4em\relax IEEE, 2019, pp.
  100--105.

\bibitem{thomsen2012effective}
J.~R. Thomsen, M.~L. Yiu, and C.~S. Jensen, ``Effective caching of shortest
  paths for location-based services,'' in \emph{Proceedings of the 2012 ACM
  SIGMOD International Conference on Management of Data}.\hskip 1em plus 0.5em
  minus 0.4em\relax ACM, 2012, pp. 313--324.

\bibitem{geisberger2008contraction}
R.~Geisberger, P.~Sanders, D.~Schultes, and D.~Delling, ``Contraction
  hierarchies: Faster and simpler hierarchical routing in road networks,'' in
  \emph{International Workshop on Experimental and Efficient Algorithms}.\hskip
  1em plus 0.5em minus 0.4em\relax Springer, 2008, pp. 319--333.

\bibitem{ouyang2018hierarchy}
D.~Ouyang, L.~Qin, L.~Chang, X.~Lin, Y.~Zhang, and Q.~Zhu, ``When hierarchy
  meets 2-hop-labeling: Efficient shortest distance queries on road networks,''
  in \emph{Proceedings of the 2018 International Conference on Management of
  Data}.\hskip 1em plus 0.5em minus 0.4em\relax ACM, 2018, pp. 709--724.

\bibitem{samet2008scalable}
H.~Samet, J.~Sankaranarayanan, and H.~Alborzi, ``Scalable network distance
  browsing in spatial databases,'' in \emph{Proceedings of the 2008 ACM SIGMOD
  international conference on Management of data}.\hskip 1em plus 0.5em minus
  0.4em\relax ACM, 2008, pp. 43--54.

\bibitem{li2017minimal}
L.~Li, W.~Hua, X.~Du, and X.~Zhou, ``Minimal on-road time route scheduling on
  time-dependent graphs,'' \emph{Proceedings of the VLDB Endowment}, vol.~10,
  no.~11, pp. 1274--1285, 2017.

\bibitem{batz2009time}
G.~V. Batz, D.~Delling, P.~Sanders, and C.~Vetter, ``Time-dependent contraction
  hierarchies,'' in \emph{Proceedings of the Meeting on Algorithm Engineering
  \& Expermiments}.\hskip 1em plus 0.5em minus 0.4em\relax Society for
  Industrial and Applied Mathematics, 2009, pp. 97--105.

\bibitem{8731400}
L.~{Li}, S.~{Wang}, and X.~{Zhou}, ``Time-dependent hop labeling on road
  network,'' in \emph{2019 IEEE 35th International Conference on Data
  Engineering (ICDE)}, April 2019, pp. 902--913.

\bibitem{zhao2012efficient}
X.~Zhao, C.~Xiao, X.~Lin, and W.~Wang, ``Efficient graph similarity joins with
  edit distance constraints,'' in \emph{2012 IEEE 28th International Conference
  on Data Engineering}.\hskip 1em plus 0.5em minus 0.4em\relax IEEE, 2012, pp.
  834--845.

\bibitem{gouda2016csi_ged}
K.~Gouda and M.~Hassaan, ``Csi\_ged: An efficient approach for graph edit
  similarity computation,'' in \emph{2016 IEEE 32nd International Conference on
  Data Engineering (ICDE)}.\hskip 1em plus 0.5em minus 0.4em\relax IEEE, 2016,
  pp. 265--276.

\bibitem{li2018efficient}
Z.~Li, X.~Jian, X.~Lian, and L.~Chen, ``An efficient probabilistic approach for
  graph similarity search,'' in \emph{2018 IEEE 34th International Conference
  on Data Engineering (ICDE)}.\hskip 1em plus 0.5em minus 0.4em\relax IEEE,
  2018, pp. 533--544.

\bibitem{chen2019efficient}
L.~Chen, Y.~Gao, Y.~Zhang, C.~S. Jensen, and B.~Zheng, ``Efficient and
  incremental clustering algorithms on star-schema heterogeneous graphs,'' in
  \emph{2019 IEEE 35th International Conference on Data Engineering
  (ICDE)}.\hskip 1em plus 0.5em minus 0.4em\relax IEEE, 2019, pp. 256--267.

\bibitem{li2018go}
L.~Li, K.~Zheng, S.~Wang, W.~Hua, and X.~Zhou, ``Go slow to go fast: minimal
  on-road time route scheduling with parking facilities using historical
  trajectory,'' \emph{The VLDB Journal—The International Journal on Very
  Large Data Bases}, vol.~27, no.~3, pp. 321--345, 2018.

\bibitem{batz2010time}
G.~V. Batz, R.~Geisberger, S.~Neubauer, and P.~Sanders, ``Time-dependent
  contraction hierarchies and approximation,'' in \emph{International Symposium
  on Experimental Algorithms}.\hskip 1em plus 0.5em minus 0.4em\relax Springer,
  2010, pp. 166--177.

\bibitem{li2020fast}
L.~Li, M.~Zhang, W.~Hua, and X.~Zhou, ``Fast query decomposition for batch
  shortest path processing in road networks,'' in \emph{2020 IEEE 36th
  International Conference on Data Engineering (ICDE)}.

\bibitem{yan2005substructure}
X.~Yan, P.~S. Yu, and J.~Han, ``Substructure similarity search in graph
  databases,'' in \emph{Proceedings of the 2005 ACM SIGMOD international
  conference on Management of data}.\hskip 1em plus 0.5em minus 0.4em\relax
  ACM, 2005, pp. 766--777.

\bibitem{zhou2009graph}
Y.~Zhou, H.~Cheng, and J.~X. Yu, ``Graph clustering based on
  structural/attribute similarities,'' \emph{Proceedings of the VLDB
  Endowment}, vol.~2, no.~1, pp. 718--729, 2009.

\bibitem{wang2010efficiently}
G.~Wang, B.~Wang, X.~Yang, and G.~Yu, ``Efficiently indexing large sparse
  graphs for similarity search,'' \emph{IEEE Transactions on Knowledge and Data
  Engineering}, vol.~24, no.~3, pp. 440--451, 2010.

\bibitem{ester1996density}
M.~Ester, H.-P. Kriegel, J.~Sander, X.~Xu \emph{et~al.}, ``A density-based
  algorithm for discovering clusters in large spatial databases with noise.''
  in \emph{Kdd}, vol.~96, no.~34, 1996, pp. 226--231.

\bibitem{gan2015dbscan}
J.~Gan and Y.~Tao, ``Dbscan revisited: mis-claim, un-fixability, and
  approximation,'' in \emph{Proceedings of the 2015 ACM SIGMOD international
  conference on management of data}.\hskip 1em plus 0.5em minus 0.4em\relax
  ACM, 2015, pp. 519--530.

\bibitem{defays1977efficient}
D.~Defays, ``An efficient algorithm for a complete link method,'' \emph{The
  Computer Journal}, vol.~20, no.~4, pp. 364--366, 1977.

\end{thebibliography}
	
\end{document}